\documentclass[aps,showpacs,amsmath,amssymb]{revtex4}

\usepackage{latexsym}
\usepackage{graphicx}
\usepackage{dcolumn}
\usepackage{bm}
\usepackage{hyperref}

\newcommand{\beq}{\begin{equation}}
\newcommand{\eeq}{\end{equation}}
\newcommand{\beqa}{\begin{eqnarray}}
\newcommand{\eeqa}{\end{eqnarray}}
\newcommand{\beqar}{\begin{eqnarray*}}
\newcommand{\eeqar}{\end{eqnarray*}}
\newcommand{\bra}[1]{\mbox{$\left\langle{#1}\right|$}}
\newcommand{\ket}[1]{\mbox{$\left|{#1}\right\rangle$}}
\newcommand{\diracsp}[2]{\mbox{$\langle{#1}|{#2}\rangle$}}

\def\I{{\rm i}}
\def\d{{\rm d}}
\def\e{{\rm e}}

\newcounter{saveeqn}

\begin{document}

\title{Quantum mechanics in general quantum systems (I): \\ exact solution}
\author{An Min Wang}\email{anmwang@ustc.edu.cn}
\affiliation{Quantum Theory Group, Department of Modern Physics,
University of Science and Technology of China, Hefei, 230026,
P.R.China}

\begin{abstract}

After proving and using a mathematical identity, we first deduce an
expansion formula of operator binomials power. Then, starting from
our idea of combining the Feynman path integral spirit and the Dyson
series kernel, we find an explicit and general form of time
evolution operator that is a $c$-number function and a power series
of perturbation including all order approximations in the
unperturbed Hamiltonian representation. Based on it, we obtain an
exact solution of the Schr\"{o}dinger equation in general quantum
systems independent of time. In special, we write down the concrete
form of the exact solution of the Schr\"odinger equation when the
solvable part (unperturbed part) of this Hamiltonian is simply taken
as the kinetic energy term. Comparison of our exact solution with
the existed perturbation theory makes some features and significance
of our exact solution clear. The conclusions expressly indicate that
our exact solution is obviously consistent with the usual
time-independent perturbation theory at any order approximation, it
explicitly calculates out the expanding coefficients of the
unperturbed state in the non-perturbation method, and it fully
solves the recurrence equation of the expansion coefficients of
final state in the unperturbed Hamiltonian representation from a
view of time-dependent perturbation theory. At the same time, the
exact solution of the von Neumann equation is also given. Our
results can be thought of as theoretical developments of quantum
dynamics, and are helpful for understanding the dynamical behavior
and related subjects of general quantum systems in both theory and
application. Our exact solution, together with its sequence studies
on perturbation theory [An Min Wang, quant-ph/0611217] and open
system dynamics [An Min Wang, quant-ph/0601051] can be used to
establish the foundation of theoretical formulism of quantum
mechanics in general quantum systems. Further applications of our
exact solution to quantum theory can be expected.

\end{abstract}


\maketitle

\section{Introduction}\label{sec1}

One of the most important tasks of physics is to obtain the time
evolution law and form of a physical system, that is, so-called
dynamical equation and its solution. This directly relates to
physics foundation. In a quantum system, the dynamical equation is
the Schr\"{o}dinger equation \cite{seq,diracpqm} for a pure state
$\ket{\Psi(t)}$ or the von Neumann equation for a mixed state
${\rho}(t)$ \cite{vonneumann}, that is \beqa\label{firstse}
-\I\frac{\partial}{\partial
t}\ket{\Psi(t)}=H\ket{\Psi(t)},\\
\label{firstvne} \dot{\rho}(t)=-\I\left[H,\rho(t)\right].\eeqa where
$H$ is the Hamiltonian of a quantum system. However, only a few
idealized quantum systems are exactly solvable at present by using
the existed theory and methods. Therefore, it is extremely
interesting and essential important to find the exact solutions to
the Schr\"odinger equation and the von Neumann equation for
Hamiltonians of even moderate complexity.

If the Hamiltonian is assumed independent of time, an arbitrary
initial state at $t=0$ is denoted by $\ket{\Psi(0)}$ for a pure
state or $\rho(0)$ for a mixed state, then the final state at a
given time $t$ will become
\beqa\label{eq.1} \ket{\Psi(t)}=\e^{-\I H t}\ket{\Psi(0)},\\
\label{eq.2}\rho(t)=\e^{-\I H t}\rho(0)\e^{\I H t},\eeqa where
$\e^{-\I H t}$ (set $\hbar=1$ for simplicity) is called as the time
evolution operator. The above equations indicate that that the final
state can be rewritten as the form that the time evolution operator
acts on an initial state. In other words, they are, respectively, a
formal solution of the Schr\"{o}dinger equation for a pure state and
one of the von Neumann equation for a mixed state. However, the
formal solution is not the practical solution that can be applied to
concrete problems and actual calculations. Ones are interested in
the practical solution because it can really produce the physical
results and conclusions in the quantum theory. Hence, to solve the
Schr\"odinger equation and the von Neumann equation refers to find
the practical solution in most cases. It is clear that this task is
equivalently to seek for the explicit expression of the time
evolution operator.

Historically, the most famous example to study the expression of the
time evolution operator is Feynman path integral formulism
\cite{Feynman}. In fact, Dyson series is also an important example
\cite{Dyson}. From our point of view, only if the expression of the
time evolution operator is a $c$-number function, can we clearly
express the practical solution of the final state using (\ref{eq.1})
and (\ref{eq.2}), respectively, for the pure state and the mixed
state. Moreover, we realize that for a general quantum system
without the exact solution in the usual theory, the expression of
the final state in $c$-number function form can be generally
expressed by an infinite series. Hence, only if the expressing
series of the final state is a power series of perturbation, can we
inherit the advantages of perturbation theory. As is well-known, to
find the $c$-number function form of the time evolution operator is
a successful linchpin of Feynman path integral formulism, and to
expand the time evolution operator as a power series of perturbation
is a powerful headstream of Dyson series (in the interaction
picture). Consequently, our physical idea in this paper comes from
the combination of Feynman path integral spirit and Dyson series
kernel. Starting from this combined idea, we would like to find an
explicit and general expression of the time evolution operator that
it not only is a $c$-number function but also is a power series
expansion of perturbation including all order approximations in
terms of our own methods.

To express the time evolution operator as a $c$-number function
needs a representation of Hilbert space, and further to expand it as
a power series of perturbation needs to separate its unperturbed
part and perturbing part. Actually, the above problems can be
studied by using the similar methods to the usual perturbation
theory \cite{seq,diracpqm}.

In the usual perturbation theory, the key idea to research the time
evolution of general quantum systems is to split the system
Hamiltonian into two parts, that is \beq H=H_0+H_1,\eeq where the
eigenvalue problem of so-called unperturbed Hamiltonian $H_0$ is
solvable, and so-called perturbing Hamiltonian $H_1$ is the rest
part of the Hamiltonian. In other words, this splitting is chosen in
such a manner that the solutions of $H_0$ are known as
\beq\label{h0eeq} H_0\ket{\Phi^\gamma}=E_\gamma \ket{\Phi^\gamma},
\eeq where $\ket{\Phi^\gamma}$ is the eigenvector of $H_0$ and
$E_\gamma$ is the corresponding eigenvalue. Whole
$\ket{\Phi^\gamma}$, in which $\gamma$ takes over all possible
values, form a representation of the unperturbed Hamiltonian. It
must be pointed out that in the perturbation theory, the above
Hamiltonian split principle with the best solvability is not unique
in more general cases. It is possible to need to consider such an
additional condition that the selected split can remove degeneracies
as complete as possible in order to suit to apply the perturbation
theory, or to let our improved scheme of perturbation theory
\cite{My2} work well. If the remained degeneracies are allowed, it
requires that the off-diagonal element of the perturbing Hamiltonian
matrix between arbitrary two degenerate levels are always vanishing.
As an example, it has been discussed in our sequence study
\cite{My3}. Of course, if the cut-off approximation of perturbation
is necessary in our improved scheme of perturbation theory
\cite{My2}, the off-diagonal elements of $H_1$ should be small
enough compared with the diagonal elements of $H=H_0+H_1$ in the
$H_0$ representation.

As soon as we find the explicit and general expression of the time
evolution operator that is both a $c$-number function and a power
series of perturbation including all order approximations, we
definitely can obtain the exact solutions of the Schr\"odinger
equation and von Neumann equation, and so we can solve a lot of
problems in quantum theory, for example, entanglement dynamics and
open system dynamics. Of course, our solutions are called ``exact
ones" in the sense including all order approximations of
perturbation. However, our final purpose is not limit to these. We,
starting from our exact solutions, try to finally establish the
foundations of theoretical formulism of quantum mechanics in general
quantum systems. Our serial studies on perturbation theory
\cite{My2} and open system dynamics \cite{My3} indicate that we have
partially arrived at our purpose.

In this paper, the key and central task is, in the unperturbed
Hamiltonian representation, to find an explicit expression of the
time evolution operator $\e^{-\I Ht}$ that is a $c$-number function
and a power series of perturbation including all order
approximations according to our idea combining the Feynman path
integral spirit and Dyson series kernel. Our method is first to
derive out an expansion formula of operator binomials power and then
apply it to the Taylor's expansion of time evolution operator
$\e^{-\I Ht}=\e^{-\I (H_0+H_1)t}$. Moreover, after proving and using
our identity, we derive out the explicit and general form of
representation matrix of $\e^{-\I Ht}$ in the representation of the
unperturbed Hamiltonian $H_0$. Consequently, we obtain the exact
solutions of the Schr\"odinger equation and the von Neumann equation
in general quantum systems independent of time, in particular, a
concrete form of the Schr\"odinger equation solution when the
solvable part of Hamiltonian is simply taken as the kinetic energy
term in order to account for the generality and universality of our
exact solutions. Furthermore, by comparing our exact solution of the
Schr\"odinger equation with the usual perturbation theory, we reveal
their relations and show what is more and what is different. The
conclusions clearly indicate that our exact solution is obviously
consistent with the usual time-independent perturbation theory at
any order approximation, but also in our exact solution we
explicitly calculate out the expanding coefficients of unperturbed
state in Lippmann-Schwinger equation \cite{LSE} for the
non-perturbative method, and/or fully solves the recurrence equation
of the expansion coefficients of final state in the unperturbed
Hamiltonian representation from a view of time-dependent
perturbation theory. Based on the above all of reasons and results,
our exact solutions can be thought of as the theoretical
developments of quantum dynamics in general quatum systems, and they
are helpful for understanding the dynamical behavior and related
subjects of quantum systems in both theory and application. From our
point of view, our exact solutions and their sequence studies on
perturbation theory \cite{My2} and open system dynamics \cite{My3}
can be used to establish the foundation of quantum mechanics of
general quantum systems. Specially, we think that the features and
advantages of our exact solutions can not be fully revealed only by
the improved scheme of perturbation theory \cite{My2} and open
system dynamics \cite{My3}. We would like to study the more
applications to the formulation of quantum mechanics in general
quantum systems in the near future.

This paper is organized as the following: in this section, we give
an introduction; in Sec. \ref{sec2} we first propose the expansion
formula of power of operator binomials; in Sec. \ref{sec3} we derive
out an explicit and general expression of time evolution operator
after proving and using our identity; in Sec. \ref{sec4} we obtain
the exact solutions of the Schr\"odinger equation and the von
Neumann equation, and then present a concrete example when the
solvable part (unperturbed part) of Hamiltonian is taken as the
kinetic energy term; in Sec. \ref{sec5} by comparison of our exact
solution of the Schr\"odinger equation with the usual perturbation
theory, we prove their consistency and their relations, and explain
what is more and what is different.; in Sec. \ref{sec6} we summarize
our conclusions and give some discussions. Finally, we write an
appendix where the proof of our identity is presented.

\section{Expansion formula of operator binomials power}\label{sec2}

In order to obtain the explicitly exact solutions of the
Schr\"odinger equation and von Neumann equation in general quantum
systems independent of time, we need to deduce the expression of the
time evolution operator. According to our physical idea, we should
first separate the time evolution operator into two parts,
respectively, with perturbation and without perturbation. After to
split the Hamiltonian, this problem changes to the derivation of
expansion formula of operator binomials power. Without loss of
generality, we are always able to write the power of operator
binomials as two parts  \beq\label{fnd} (A+B)^n=A^n+f^n(A,B). \eeq
where $A$ and $B$ are two operators and do not commute with each
other in general. If $B$ is taken as a perturbation, the above
decomposition just splits the operator binomials power into two
parts, respectively, with perturbation and without perturbation.
Now, the key matter is how to obtain the form of $f^n(A,B)$.

It is clear that $f^n(A,B)$ is a polynomial including at least first
power of $B$ and at most $n$th power of $B$ in every term. Thus, a
general term with $l$th power of $B$ has the form $
\left(\prod_{i=1}^lA^{k_i}B\right)A^{n-l-\sum_{i=1}^l k_i}$. From
the symmetry of power of binomials we conclude that every $k_i$ take
the values from 0 to $(n-l)$, but it must keep $n-l-\sum_{i=1}^l
k_i\geq 0$. So we have \beqa\label{fne}
f^n(A,B)\!\!&=&\!\!\sum_{l=1}^n\!\!\sum_{\stackrel{\scriptstyle
k_1,\cdots,k_l=0}{\sum_{i=1}^l k_i+l\leq
n}}^{n-l}\!\!\!\left(\prod_{i=1}^lA^{k_i}B\right)\!A^{n-l-\sum_{i=1}^l
k_i}\\
&=&\label{fne1}\sum_{l=1}^n\sum_{k_1,\cdots,k_l=0}^{n-l}
\left(\prod_{i=1}^lA^{k_i}B\right)A^{n-l-\sum_{i=1}^l
k_i}\theta\left(n-l-\sum_{i=1}^l k_i\right). \eeqa where $\theta(x)$
is a step function, that is, $\theta(x)=1$ if $x\geq 0$, and
$\theta(x)=0$ if $x<0$. Obviously based on above definition, we
easily verify \beq f^1(A,B)=B,\quad f^2(A,B)=AB+B(A+B). \eeq They
imply that the expression (\ref{fne}) is correct for $n=1,2$.

Now we use the mathematical induction to prove the expression
(\ref{fne}) of $f^n(A,B)$, that is, let us assume that it is valid
for a given $n$, and then prove that it is also valid for $n+1$.
Denoting ${\mathcal{F}}^{n+1}(A,B)$ with the form of expression
(\ref{fne1}) where $n$ is replaced by $n+1$, and we extract the part
of $l=1$ in its finite summation for $l$ \beqa
\mathcal{F}^{n+1}(A,B)&=&\sum_{k_1=0}^{n}A^{k_1}BA^{n-k_1}+\sum_{l=2}^{n+1}
\sum_{k_1,\cdots,k_l=0}^{(n+1)-l}\left(\prod_{i=1}^lA^{k_i}B\right)A^{(n+1)-l-\sum_{i=1}^l
k_i}\theta\left((n+1)-l-\sum_{i=1}^l k_i\right).\eeqa We extract the
terms $k_1=0$ in the first and second summations, and again replace
$k_1$ by $k_1-1$ in the summation (the summations for $k_i$ from 1
to $(n+1)-l$ change as one from 0 to $n-l$), the result is \beqa
\mathcal{F}^{n+1}(A,B)&=&B A^n+\sum_{k_1=1}^{n}A^{k_1}BA^{n-k_1}+
B\sum_{l=2}^{n+1}
\sum_{k_2,\cdots,k_l=0}^{(n+1)-l}\left(\prod_{i=2}^lA^{k_i}B\right)
A^{(n+1)-l-\sum_{i=2}^l
k_i}\theta\left((n+1)-l-\sum_{i=2}^l k_i\right) \nonumber\\
& & +\sum_{l=2}^{n+1}\sum_{k_1=1}^{(n+1)-l}
\sum_{k_2,\cdots,k_l=0}^{(n+1)-l}\left(\prod_{i=1}^lA^{k_i}B\right)
A^{(n+1)-l-\sum_{i=1}^l k_i}\theta\left((n+1)-l-\sum_{i=1}^l
k_i\right), \eeqa furthermore \beqa \mathcal{F}^{n+1}(A,B)&=&B
A^n+A\sum_{k_1=0}^{n-1}A^{k_1}BA^{n-1-k_1}+ B\sum_{l=2}^{n+1}
\sum_{k_2,\cdots,k_l=0}^{(n+1)-l}\left(\prod_{i=2}^lA^{k_i}B\right)\nonumber\\
& &\times A^{(n+1)-l-\sum_{i=2}^l
k_i}\theta\left((n+1)-l-\sum_{i=2}^l k_i\right) \nonumber\\
& & +A\sum_{l=2}^{n+1}\sum_{k_1=0}^{n-l}
\sum_{k_2,\cdots,k_l=0}^{(n+1)-l}\left(\prod_{i=1}^lA^{k_i}B\right)A^{n-l-\sum_{i=1}^l
k_i}\theta\left(n-l-\sum_{i=1}^l k_i\right).\label{fneproof}\eeqa
Considering the third term in above expression, we change the dummy
index $\{k_2,k_3,\cdots,k_l\}$ into $\{k_1,k_2,\cdots,k_{l-1}\}$,
rewrite $(n+1)-l$ as $n-(l-1)$, and finally replace $l$ by $l-1$ in
the summation (the summation for $l$ from 2 to $n+1$ changes as one
from 1 to $n$), we obtain \beqa & &B\sum_{l=2}^{n+1}
\sum_{k_2,\cdots,k_l=0}^{(n+1)-l}\left(\prod_{i=2}^lA^{k_i}B\right)A^{(n+1)-l-\sum_{i=2}^l
k_i}\theta\left((n+1)-l-\sum_{i=2}^l k_i\right)\nonumber\\
& &\quad = B\sum_{l=2}^{n+1}
\sum_{k_1,\cdots,k_{l-1}=0}^{n-(l-1)}\left(\prod_{i=1}^{l-1}A^{k_i}B\right)A^{n-(l-1)-\sum_{i=1}^{l-1}
k_i}\theta\left(n-(l-1)-\sum_{i=1}^{l-1} k_i\right)\nonumber\\
& &\quad = B\sum_{l=1}^{n}
\sum_{k_1,\cdots,k_{l}=0}^{n-l}\left(\prod_{i=1}^{l}A^{k_i}B\right)A^{n-l-\sum_{i=1}^{l}
k_i}\theta\left(n-l-\sum_{i=1}^{l} k_i\right)\nonumber\\ & &\quad=
Bf^n(A,B)\label{fnet3},\eeqa where we have used the expression
(\ref{fne1}) of $f^n(A,B)$. Because of the step function
$\theta\left(n-l-\sum_{i=1}^l k_i\right)$ in the fourth term of the
expression (\ref{fneproof}), the upper bound of summation for $l$ is
abated to $n$, and the upper bound of summation for
$k_2,k_3,\cdots,k_l$ is abated to $n-l$. Then, merging the fourth
term and the second term in Eq.(\ref{fneproof}) gives \beqa &
&A\sum_{k_1=0}^{n-1}A^{k_1}BA^{n-1-k_1}+A\sum_{l=2}^{n+1}\sum_{k_1=0}^{n-l}
\sum_{k_2,\cdots,k_l=0}^{(n+1)-l}\left(\prod_{i=1}^lA^{k_i}B\right)
A^{n-l-\sum_{i=1}^l
k_i}\theta\left(n-l-\sum_{i=1}^l k_i\right)\nonumber\\
& &\quad =A\sum_{k_1=0}^{n-1}A^{k_1}BA^{n-1-k_1}+A\sum_{l=2}^{n}
\sum_{k_1,k_2,\cdots,k_l=0}^{n-l}\left(\prod_{i=1}^lA^{k_i}B\right)
A^{n-l-\sum_{i=1}^l
k_i}\theta\left(n-l-\sum_{i=1}^l k_i\right)\nonumber\\
& &\quad =A\sum_{l=1}^{n}
\sum_{k_1,k_2,\cdots,k_l=0}^{n-l}\left(\prod_{i=1}^lA^{k_i}B\right)A^{n-l-\sum_{i=1}^l
k_i}\theta\left(n-l-\sum_{i=1}^l k_i\right)\nonumber\\
& &\qquad =A f^n(A,B)\label{fnet24},\eeqa where we have used again
the expression (\ref{fne1}) of $f^n(A,B)$.

Substituting (\ref{fnet3}) and (\ref{fnet24}) into (\ref{fneproof}),
immediately leads to the following result \beq
\label{fnadd1org}\mathcal{F}^{n+1}=B A^n +(A+B)f^n(A,B).\eeq

Note that
\beqa (A+B)^{n+1}&=&(A+B)(A+B)^n\nonumber\\
&=&A^{n+1}+B A^{n}+(A+B)f^{n}(A,B),\eeqa we have the relation \beq
\label{fnadd1}f^{n+1}(A,B)=B A^{n}+(A+B)f^{n}(A,B).\eeq Therefore,
in terms of Eqs.(\ref{fnadd1org}) and (\ref{fnadd1}) we have
finished our proof that  $f^n(A,B)$ has the expression (\ref{fne})
or (\ref{fne1}) for any $n$.

\section{Expression of the time evolution operator}\label{sec3}

Now we investigate the expression of the time evolution operator by
means of above expansion formula of power of operator binomials,
that is, we write \beq e^{-\I H t}=\sum_{n=0}^\infty\frac{(-\I
t)^n}{n!}(H_0+H_1)^n=\e^{-\I H_0 t}+\sum_{n=0}^\infty\frac{(-\I
t)^n}{n!}f^n(H_0,H_1). \eeq In above equation, inserting the
complete relation $
\sum_{\gamma}\ket{\Phi^{\gamma}}\bra{\Phi^{\gamma}}=1 $ before every
$H_0^{k_i}$, and using the eigen equation of $H_0$ (\ref{h0eeq}), it
is easy to see that \beqa \label{fk}
f^n(H_0,H_1)&=&\sum_{l=1}^n\sum_{\gamma_1,\cdots,\gamma_{l+1}}
\sum_{\stackrel{\scriptstyle k_1,\cdots,k_l=0}{\sum_{i=1}^l
k_i+l\leq n}}^{n-l}\left[\prod_{i=1}^l E_{\gamma_i}^{k_i}\right]
E_{\gamma_{l+1}}^{n-\sum_{i=1}^lk_i-l}
\left[\prod_{i=1}^lH_1^{\gamma_i\gamma_{i+1}}\right]
\ket{\Phi^{\gamma_1}}\bra{\Phi^{\gamma_{l+1}}}\nonumber\\
&=&\sum_{l=1}^n\sum_{\gamma_1,\cdots,\gamma_{l+1}}C_l^n(E[\gamma,l])
\left[\prod_{i=1}^lH_1^{\gamma_i\gamma_{i+1}}\right]
\ket{\Phi^{\gamma_1}}\bra{\Phi^{\gamma_{l+1}}}, \label{fkc}\eeqa
where $H_1^{\gamma_i\gamma_{i+1}}
=\bra{\Phi^{\gamma_{i}}}H_1\ket{\Phi^{\gamma_{i+1}}}$, $E[\gamma,l]$
is a vector with $l+1$ components denoted by \beq E[\gamma,l]
=\{E_{\gamma_1},E_{\gamma_2},
\cdots,E_{\gamma_l},E_{\gamma_{l+1}}\}\eeq and we introduce the
definition of $C^n_l(E[\gamma,l])$ ($l\geq 1$) as the following
\beq\label{cd1} C_l^n(E[\gamma,l])=\sum_{\stackrel{\scriptstyle
k_1,\cdots,k_l=0}{\sum_{i=1}^l k_i+l\leq
n}}^{n-l}\!\!\!\left[\prod_{i=1}^l E_{\gamma_i}^{k_i}\right]
E_{\gamma_{l+1}}^{n-\sum_{i=1}^lk_i-l}.\eeq The expression
(\ref{fk}) and the expression of the time evolution operator depend
on it has incarnated our physical idea, that is, it is both a
$c$-number function and a power series of perturbation. Our further
task is to find the explicit expression of $C^n_l(E[\gamma,l])$.

Before to find $C^n_l(E[\gamma,l])$, let us mention the cases when
energy level degenerations happen, that is, some same values appear
in the eigen spectrum
$\mathfrak{E}=\{E_{\gamma}|\gamma=1,2,\cdots\}$, the eigen spectrum
is re-denoted by $\mathfrak{E}=\{E_{\gamma
a_\gamma}|\gamma=1,2,\cdots, a_\gamma=1,\cdots,m_\gamma\}$ and the
numbers of original $\gamma$ is different from the numbers of new
$\gamma$. Here, $m_\gamma$ is called degeneracy for a given
$\gamma$, which means the number of the same eigenvalues
($m_\gamma$-fold degeneracy). In special, when $m_\gamma=1$, the
energy level $E_{\gamma a_\gamma}$ is not degenerate. It is clear
that the above expression (\ref{fkc}) has only a little change:
\beqa f^n(H_0,H_1)
&=&\sum_{l=1}^n\sum_{\gamma_1,\cdots,\gamma_{l+1}}
\sum_{a_{\gamma_1},\cdots,a_{\gamma_{l+1}}}C_l^n(E[\gamma,l])
\left[\prod_{i=1}^lH_1^{\gamma_ia_{\gamma_i},\gamma_{i+1}a_{\gamma_{i+1}}}\right]
\ket{\Phi^{\gamma_1
a_{\gamma_1}}}\bra{\Phi^{\gamma_{l+1}a_{\gamma_{l+1}}}}, \hskip
1.0cm\eeqa which has no obvious influence on our following
derivation and proof. For simplicity, we do not consider the
degenerate cases in this section since no new idea and skill will be
needed here.

In order to derive out an explicit and useful expression of
$C^n_l(E[\gamma,l])$, we first change the dummy index
$k_l\rightarrow n-l-\sum_{i=1}^lk_i$ in Eq.(\ref{fne1}). Note that
at the same time in spite of $\theta(n-l-\sum_{i=1}^lk_i)$ changes
as $\theta(k_l)$, but a hiding factor $\theta(k_l)$ becomes
$\theta(n-l-\sum_{i=1}^lk_i)$. Thus, we can rewrite
\beqa\label{fennew}
f^n(A,B)&=&\sum_{l=1}^n\sum_{k_1,\cdots,k_l=0}^{n-l}
\left(\prod_{i=1}^{l-1}A^{k_i}B\right)A^{n-l-\sum_{i=1}^l k_i}B
A^{k_l}\theta\left(n-l-\sum_{i=1}^l k_i\right). \eeqa In fact, this
new expression of $f^n(A,B)$ is a result of the symmetry of power of
binomials for its every factor. Similarly, in terms of the above
method to obtain the definition of $C^n_l(E[\gamma,l])$ (\ref{cd1}),
we have its new definition (where $k_l$ is replaced by $k$) \beqa
C_l^n(E[\gamma,l])
&=&\sum_{k=0}^{n-l}\left\{\sum_{k_1,\cdots,k_{l-1}=0}^{n-l}\left[\prod_{i=1}^{l-1}
E_{\gamma_i}^{k_i}\right]E_{\gamma_l}^{(n-k)-\sum_{i=1}^{l-1}k_i-l}
\theta\left((n-k)-l-\sum_{i=1}^{l-1}k_i\right)\right\}E^k_{\gamma_{l+1}}.
\eeqa Because $\theta\left((n-k)-l-\sum_{i=1}^{l-1}k_i\right)$
abates the upper bound of summation for $k_i (i=1,2,\cdots,l-1)$
from $(n-l)$ to $(n-k-1)-(l-1)$, we obtain the recurrence equation
\beqa C_l^n(E[\gamma,l])
&=&\sum_{k=0}^{n-l}\left\{\sum_{k_1,\cdots,k_{l-1}=0}^{(n-k-1)-(l-1)}\left[\prod_{i=1}^{l-1}
E_{\gamma_i}^{k_i}\right]E_{\gamma_l}^{(n-k-1)-\sum_{i=1}^{l-1}k_i-(l-1)}\right.\nonumber\\
& &\left. \times
\theta\left((n-k-1)-(l-1)-\sum_{i=1}^{l-1}k_i\right)\right\}E^k_{\gamma_{l+1}}\nonumber\\
&=&\sum_{k=0}^{n-l}C_{l-1}^{n-k-1}(E[\gamma,l-1])
E_{\gamma_{l+1}}^k. \label{creq} \eeqa

In particular, when $l=1$, from the definition of
$C^n_l(E(\gamma,1))$ it follows that \beq
C^n_1(E(\gamma,1))=\sum_{k_1=0}^{n-1}E_{\gamma_1}^{k_1}E_{\gamma_2}^{n-1-k_1}
=E_{\gamma_2}^{n-1}\sum_{k_1=0}^{n-1}\left(\frac{E_{\gamma_1}}{E_{\gamma_2}}\right)^{k_1}.\eeq
By means of the summation formula of a geometric series, we find
that \beq
C^n_1(E(\gamma,1))=\frac{E_{\gamma_1}^n}{E_{\gamma_1}-E_{\gamma_2}}
-\frac{E_{\gamma_2}^n}{E_{\gamma_1}-E_{\gamma_2}}.\eeq Based on the
recurrence equation (\ref{creq}), we have \beqa
C^n_2(E(\gamma,2))&=&\sum_{k=0}^{n-2}\left(\frac{E_{\gamma_1}^{n-k-1}}{E_{\gamma_1}-E_{\gamma_2}}
-\frac{E_{\gamma_2}^{n-k-1}}{E_{\gamma_1}-E_{\gamma_2}}\right)E_{\gamma_3}^k\nonumber\\
&=&
\sum_{i=1}^2\frac{(-1)^{i-1}E_{\gamma_i}^{n-1}}{E_{\gamma_1}-E_{\gamma_2}}
\sum_{k=0}^{n-2}\left(\frac{E_{\gamma_3}}{E_{\gamma_i}}\right)^{k}\nonumber\\
&=&\sum_{i=1}^2\frac{(-1)^{i-1}(E_{\gamma_i}^n-E_{\gamma_i}E_{\gamma_3}^{n-1})}
{(E_{\gamma_1}-E_{\gamma_2})(E_{\gamma_i}-E_{\gamma_3})}\nonumber\\
&=&\frac{E_{\gamma_1}^n}{(E_{\gamma_1}-E_{\gamma_2})(E_{\gamma_1}-E_{\gamma_3})}
-\frac{E_{\gamma_2}^n}{(E_{\gamma_1}-E_{\gamma_2})(E_{\gamma_2}-E_{\gamma_3})}\nonumber\\
&
&+\frac{E_{\gamma_3}^n}{(E_{\gamma_1}-E_{\gamma_3})(E_{\gamma_2}-E_{\gamma_3})}.\eeqa
In the above calculations, the last step is important. In fact, in
order to obtain the concrete expression of $C_l^n$ ($l$ and $n$ are
both positive integers), we need our identity \beq
\label{myi}\sum_{i=1}^{l+1} (-1)^{i-1}
\frac{E_{\gamma_i}^K}{d_i(E[\gamma,l])}
=\left\{\begin{array}{c l}0 &\quad (\mbox{If $0\leq K<l$})\\[8pt] 1 &\quad (\mbox{If
$K=l$})\end{array}\right.. \eeq It is proved in Appendix A in
detail. The denominators $d_i(E[\gamma,l])$ in above identity are
defined by \beqa
d_1(E[\gamma,l])&=&\prod_{i=1}^{l}\left(E_{\gamma_{1}}
-E_{\gamma_{i+1}}\right),\\
 d_i(E[\gamma,l])&=&
\prod_{j=1}^{i-1}\left(E_{\gamma_{j}}
-E_{\gamma_{i}}\right)\!\!\!\prod_{k=i+1}^{l+1}\left(E_{\gamma_{i}}
-E_{\gamma_{k}}\right),\\[-3pt] d_{l+1}(E[\gamma,l])
&=&\prod_{i=1}^{l}\left(E_{\gamma_{i}}-E_{\gamma_{l+1}}\right),\eeqa
where $2\leq i \leq l$. Then, using the recurrence equation
(\ref{creq}) and our identity (\ref{myi}), we obtain \beq\label{cne}
C^n_l(E[\gamma,l])=\sum_{i=1}^{l+1} (-1)^{i-1}
\frac{E_{\gamma_i}^n}{d_i(E[\gamma,l])}. \eeq Here, the mathematical
induction shows its power again. If the expression (\ref{cne}) is
correct for a given $n$, for example $n=1,2$, then for $n+1$ from
the recurrence equation (\ref{creq}) it follows that \beqa
C^{n+1}_l(E[\gamma,l])\!\!&=&\!\!\sum_{k=0}^{n+1-l}\left(\sum_{i=1}^{l}
(-1)^{i-1}
\frac{E_{\gamma_i}^{n-k}}{d_i(E[\gamma,l-1])}\right)E_{\gamma_{l+1}}^k\nonumber\\
&=&
\sum_{i=1}^{l}\frac{(-1)^{i-1}E_{\gamma_i}^{n}}{d_i(E[\gamma,l-1])}\sum_{k=0}^{n+1-l}
\left(\frac{E_{\gamma_{l+1}}}{E_{\gamma_{i}}}\right)^k\nonumber\\
&=&\sum_{i=1}^{l}\frac{(-1)^{i-1}(E_{\gamma_i}^{n+1}-E_{\gamma_i}^{l-1}E_{\gamma_{l+1}}^{n+2-l})}
{d_i(E[\gamma,l-1])(E_{\gamma_i}-E_{\gamma_{l+1}})}.\label{cnproof}\eeqa
Because that
$d_i(E[\gamma,l-1])(E_{\gamma_i}-E_{\gamma_{l+1}})=d_i(E[\gamma,l])$
($i\leq l+1$) and our identity (\ref{myi}), it is easy to see\beq
\sum_{i=1}^{l}\frac{(-1)^{i-1}E_{\gamma_i}^{l-1}}
{d_i(E[\gamma,l])}= -(-1)^l \frac{E_{\gamma_{l+1}}^{l-1}}
{d_{l+1}(E[\gamma,l])}. \eeq Substitute it into Eq.(\ref{cnproof})
yields \beq C^{n+1}_l(E[\gamma,l])=\sum_{i=1}^{l+1} (-1)^{i-1}
\frac{E_{\gamma_i}^{n+1}}{d_i(E[\gamma,l])}. \eeq That is, we have
proved that the expression (\ref{cne}) of $C^n_l(E[\gamma,l])$ is
valid for any $n$.

It must be emphasized that since the summation is over all of values
of the energy eigenvalues, there are apparent divergences in the
expression of $C^{n}_l(E[\gamma,l])$. Here, ``apparent" refers to an
unture thing, that is, the apparent divergences are not real
singularities and they can be eliminated by mathematical and/or
physical methods. Moreover, the same problem appears if with
degeneracy. Therefore, we need to understand above expressions in
the sense of limitations. For instance, for $
C^n_1(E[\gamma,1])=\left(E_{\gamma_{1}}^{n}
-E_{\gamma_{2}}^{n}\right)/\left({E_{\gamma_{1}}
-E_{\gamma_{2}}}\right)$, we have the expression $
\lim_{E_{\gamma_{2}} \rightarrow E_{\gamma_{1}}}C_1^n(E[\gamma,1])
=\left(\delta_{\gamma_1\gamma_2}
+\Theta(\left|\gamma_2-\gamma_1\right|)\delta_{E_{\gamma_{1}}
E_{\gamma_{2}}}\right)n E_{\gamma_{1}}^{n-1}$, where the step
function $\Theta(x)=1$, if $x>0$, and $\Theta(x)=0$, if $x\leq 0$.
In other words, the apparent divergences here are not real
divergences and can be eliminated by finding the correct
limitations. There is no theoretical problem when we formally keep
the apparent divergences in the above expression. Actually, in our
recent work \cite{My2}, we successfully eliminate all the apparent
divergences.

Because of our identity (\ref{myi}), the summation to $l$ in
Eq.(\ref{fkc}) can be extended to $\infty$. Thus, the expression of
the time evolution operator is changed to a summation according to
the order (or power) of the perturbing Hamiltonian $H_1$ as follows
\beqa \label{mk} \bra{\Phi^\gamma}\e^{-\I H
t}\ket{\Phi^{\gamma^\prime}}= \e^{-\I E_{\gamma}
t}\delta_{\gamma\gamma^\prime} + \sum_{l=1}^\infty
\sum_{\gamma_1,\cdots,\gamma_{l+1}}\left[
\sum_{i=1}^{l+1}(-1)^{i-1}\frac{\e^{-\I E_{\gamma_i}
t}}{d_i(E[\gamma,l])}\right]
\prod_{j=1}^{l}H_1^{\gamma_j\gamma_{j+1}}
\delta_{\gamma_1\gamma}\delta_{\gamma_{l+1}\gamma^\prime}.\eeqa When
there is degeneracy, it becomes \beqa \label{mkd}
\bra{\Phi^\gamma}\e^{-\I H t}\ket{\Phi^{\gamma^\prime}}&=& \e^{-\I
E_{\gamma} t}\delta_{\gamma\gamma^\prime} + \sum_{l=1}^\infty
\sum_{\gamma_1,\cdots,\gamma_{l+1}}\sum_{a_{\gamma_1},\cdots,a_{\gamma_{l+1}}}\left[
\sum_{i=1}^{l+1}(-1)^{i-1}\frac{\e^{-\I E_{\gamma_i}
t}}{d_i(E[\gamma,l])}\right]\nonumber\\
& &\times \prod_{j=1}^{l}H_1^{\gamma_j
a_{\gamma_j},\gamma_{j+1}a_{\gamma_{j+1}}}
\delta_{\gamma_1\gamma}\delta_{a_{\gamma_1}a_{\gamma}}
\delta_{\gamma_{l+1}\gamma^\prime}\delta_{a_{\gamma_{l+1}}a_{\gamma^\prime}}\\
&=& \sum_{l=0}^\infty A_l^{\gamma v,\gamma^\prime
v^\prime}(t)=\sum_{l=0}^\infty\bra{\Phi^{\gamma
}}\mathcal{A}_l(t)\ket{\Phi^{\gamma^\prime}},\eeqa that is,
\beqa\label{eteo} \e^{-\I H_{tot} t}&=&\sum_{l=0}^\infty
\mathcal{A}_l(t)=\sum_{l=0}^\infty\sum_{\gamma,\gamma^\prime}
A_l^{\gamma,\gamma^\prime}(t)\ket{{\Phi}^{\gamma}}\bra{{\Phi}^{\gamma^\prime}}\eeqa
where \beqa\label{Aldefinition}
A_0^{\gamma,\gamma^\prime}(t)&=&\e^{-\I E_{\gamma}
t}\delta_{\gamma\gamma^\prime}\\
A_l^{\gamma\gamma^\prime}(t)&=&\sum_{\gamma_1,\cdots,\gamma_{l+1}}\left[
\sum_{i=1}^{l+1}(-1)^{i-1}\frac{\e^{-\I E_{\gamma_i}
t}}{d_i(E[\gamma,l])}\right]\left[
\prod_{j=1}^{l}H_1^{\gamma_j\gamma_{j+1}}\right]
\delta_{\gamma_1\gamma}\delta_{\gamma_{l+1}\gamma^\prime}.\eeqa

It is clear that this propagator has the closed time evolution
factors. In special, it is a $c$-number function just like the
Feynman path integral \cite{Feynman}, and it is a power series of
the perturbing Hamiltonian just like Dyson series in interaction
picture \cite{Dyson}. Therefore it has, at the same time, the
advantages that exist in the Feynman path integral famulism and
Feynman diagram (Dyson) expansion formula .

\section{Solution of the Schr\"odinger equation and the von Neumann equation}\label{sec4}

Substituting the expression of the time evolution operator
(\ref{mk}) into Eq.(\ref{eq.1}), we obtain immediately our explicit
form of time evolution of an arbitrary initial state in a general
quantum system \beqa \label{ouress}
\ket{\Psi(t)}&=&\sum_{l=0}^\infty
\mathcal{A}_l(t)\ket{\Psi(0)}=\sum_{l=0}^\infty\sum_{\gamma,\gamma^\prime}
A_l^{\gamma\gamma^\prime}(t)
\left[\diracsp{\Phi^{\gamma^\prime}}{\Psi(0)}\right]\ket{\Phi^\gamma},
\eeqa In form, this solution is also suitable for the degenerate
cases, but more apparent divergences will appear. It must be
emphasized that these apparent divergences can be easily eliminated
by the limitation calculations. To save space, we delay to discuss
the disposal of degenerate cases and the elimination of apparent
divergences to our sequence study \cite{My2}.

Similarly, the exact solution of the von Neumann equation for mixed
states can be obtained by using our general and explicit expression
of the time evolution operator (\ref{eteo}). From the formal
solution of the von Neumann equation \beq \rho(t)=\e^{-\I H
t}\rho(0)\e^{\I Ht},\eeq it immediately follows that \beqa
\rho(t)&=&\sum_{k,l=0}^\infty
\mathcal{A}_k(t)\rho(0)\mathcal{A}_l(-t)=\sum_{k,l=0}^\infty
\mathcal{A}_k(t)\rho(0)\mathcal{A}^\dagger_l(t)\\
&=&\sum_{k,l=0}^\infty\sum_{\gamma,\gamma^\prime}\sum_{\beta,\beta^\prime}
A_k^{\gamma\beta}(t)\rho^{\beta\beta^\prime}(0)
A_l^{\beta^\prime\gamma^\prime}(-t)\ket{\Phi^\gamma}\bra{\Phi^{\gamma^\prime}}\eeqa

Since we have obtained the exact expression of the time evolution
operator, it is direct to extend our conclusions from the pure state
to the mixed state. For simplicity, we only study the cases of pure
states in the following, that is, our exact solution of the
Schr\"dinger equation.

Note that the linearity of the evolution operator and the
completeness of eigenvector set of $H_0$, we can simply set the
initial state as an eigenvector of $H_0$. Up to now, everything is
exact and no any approximation enters. Therefore, it is an exact
solution despite that its form is an infinity series. In other
words, it exactly includes all order approximations of $H_1$. Only
when $H_1$ is taken as a perturbation, can it be cut off based on
the needed precision. From a view of formalized theory, it is
explicit and general, but is not compact. Moreover, if the
convergence is guaranteed, it is strict since its general term is
known. Usually, to a practical purpose, if only including the finite
(often low) order approximation of $H_1$, above expression is cut
off to the finite terms and becomes a perturbed solution.

Actually, since our solution includes the contributions from all
order approximations of the rest part of Hamiltonian except for the
solvable part $H_0$, or all order approximations of the perturbative
part $H_1$ of Hamiltonian, it is not very important whether $H_1$ is
(relatively) large or small when compared with $H_0$. In principle,
for a general quantum system with the normal form Hamiltonian, we
can always write down \beq \label{Hnf} H=\frac{\hat{\bm{k}}^2}{2m}+V
.\eeq If we take the solvable kinetic energy term as
$H_0=\displaystyle\frac{\hat{\bm{k}}^2}{2m}$ and the potential
energy part as $H_1=V$, our method is applicable to such quantum
system. It should be pointed out that we study this concrete example
in order to account for the generality and universality of our exact
solution.

It is clear that \beq \frac{\hat{\bm{k}}^2}{2m}\ket{\bm{k}}
=\frac{\bm{k}^2}{2m}\ket{\bm{k}}=E_{\bm{k}}\ket{\bm{k}},\quad
E_{\bm{k}}=\frac{\bm{k}^2}{2m}. \eeq And denoting \beqa
\diracsp{\bm{x}}{\bm{k}}&=&\frac{1}{L^{3/2}}\;\e^{\I\bm{k}\cdot\bm{x}},\\
\psi_{\bm{k}}(0)&=& \diracsp{\bm{k}}{\Psi(0)},\\
\Psi(\bm{x},t)&=&\diracsp{\bm{x}}{\Psi(t)},\eeqa we obtain the final
state as \beqa \label{ffs1} \Psi(\bm{x},t)&=&\sum_{\bm{k}}
\psi_{\bm{k}}(0)\frac{\e^{\I\left(\bm{k}\cdot\bm{x}-E_{\bm{k}}t\right)}}{L^{3/2}}+
\sum_{l=1}^\infty
\sum_{\bm{k},\bm{k}^\prime}\sum_{\bm{k}_1,\cdots,\bm{k}_{l+1}}
\psi_{\bm{k}^\prime}(0)
\left[\prod_{j=1}^{l}V^{\bm{k}_j\bm{k}_{j+1}}\right]
\delta_{\bm{k}_1\bm{k}}\delta_{\bm{k}_{l+1}\bm{k}^\prime}\nonumber\\
& &\times\left[ \sum_{i=1}^{l+1}\frac{(-1)^{i-1}}{d_i(E[k,l])}\;
\frac{1}{L^{3/2}}\;\e^{\I\left(\bm{k}\cdot\bm{x}-E_{\bm{k}_i}t\right)}
\right]\\ \label{ffs2} &=&\sum_{\bm{k}}
\psi_{\bm{k}}(0)\frac{\e^{\I\left(\bm{k}\cdot\bm{x}-E_{\bm{k}}t\right)}}{L^{3/2}}+
\sum_{l=1}^\infty
\sum_{\bm{k},\bm{k}^\prime}\sum_{\bm{k}_1,\cdots,\bm{k}_{l+1}}
\psi_{\bm{k}^\prime}(0)
\left[\prod_{j=1}^{l}\mathcal{V}(\bm{k}_j-\bm{k}_{j+1})\right]
\delta_{\bm{k}_1\bm{k}}\delta_{\bm{k}_{l+1}\bm{k}^\prime}\nonumber\\
& &\times\left[ \sum_{i=1}^{l+1}\frac{(-1)^{i-1}}{d_i(E[k,l])}\;
\frac{1}{L^{3/2}}\;\e^{\I\left(\bm{k}\cdot\bm{x}-E_{\bm{k}_i}t\right)}
\right].\eeqa In the second equal mark, we have used the fact that
$V=V(\bm{x})$ usually, thus \beqa
V^{\bm{k}_j\bm{k}_{j+1}}&=& \bra{\bm{k_j}}V(\bm{x})\ket{\bm{k}_{j+1}}\nonumber\\
&=&\int^\infty_{-\infty}\d^3x_j\d^3x_{j+1}
\diracsp{\bm{k_j}}{\bm{x}_j}\bra{\bm{x}_j}V(\bm{x})\ket{\bm{x}_{j+1}}
\diracsp{\bm{x}_{j+1}}{\bm{k_{j+1}}}\nonumber\\
&=&\frac{1}{L^3}\int^\infty_{-\infty}\d^3x_j\d^3x_{j+1}
\e^{-\I\bm{k}_j\cdot\bm{x}_j}V(\bm{x}_{j+1})\delta^3(\bm{x}_j-\bm{x}_{j+1})
\e^{\I\bm{k}_{j+1}\cdot\bm{x}_{+1}j}\nonumber \\
&=&\frac{1}{L^3}\int^\infty_{-\infty}\d^3x
V(\bm{x})\e^{-\I(\bm{k_{j}}-\bm{k_{j+1}})\cdot\bm{x}}\nonumber\\
 &=&\mathcal{V}(\bm{k}_{j}-\bm{k}_{j+1}). \eeqa It is clear
 that $\mathcal{V}(\bm{k})$ is the Fourier transformation of
 $V(\bm{x})$.
Eq.(\ref{ffs1}) or (\ref{ffs2}) is a concrete form of our solution
(\ref{ouress}) of Schr\"odinger equation when the solvable part of
Hamiltonian is taken as the kinetic energy term. Therefore, in
principle, if the Fourier transformation of $V(\bm{x})$ can be
found, the evolution of the arbitrarily initial state with time can
be obtained. It must be emphasized that it is often that the
solvable part of Hamiltonian will be not only $T=\displaystyle
\frac{\hat{\bm{k}}^2}{2m}$ for the practical purposes.

Furthermore, we can derive out the propagator \beqa
\bra{\bm{x}}\e^{-\I
Ht}\ket{\bm{x}^\prime}&=&\sum_{\bm{k},\bm{k}^\prime}\diracsp{\bm{x}}{\bm{k}}\bra{\bm{k}}\e^{-\I
Ht}\ket{\bm{k}^\prime}\diracsp{\bm{k}^\prime}{\bm{x}^\prime}\nonumber\\
&=& \sum_{\bm{k}}\frac{1}{L^3}\e^{\I\bm{k}(\bm{x}-\bm{x}^\prime)-\I
E_{\bm{k}}t}+\sum_{l=1}^\infty
\sum_{\bm{k},\bm{k}^\prime}\sum_{\bm{k}_1,\cdots,\bm{k}_{l+1}}
\left[\prod_{j=1}^{l}V^{\bm{k}_j\bm{k}_{j+1}}\right]
\delta_{\bm{k}_1\bm{k}}\delta_{\bm{k}_{l+1}\bm{k}^\prime}\nonumber\\
& &\times\left[ \sum_{i=1}^{l+1}\frac{(-1)^{i-1}}{d_i(E[k,l])}\;
\frac{1}{L^{3/2}}\;\e^{-\I E_{\bm{k}_i}t}
\right]\frac{1}{L^3}\e^{\I\bm{k}\cdot\bm{x}-\I\bm{k}^\prime\cdot\bm{x}^\prime}.
\eeqa It is different from the known propagator in the expressive
form, but they should be equivalent in theory because we still use
the usual quantum mechanics principle. However, what is more and
what is new in our exact solution? In the following, and in our
serial studies \cite{My2,My3}, we will see them.

\section{Comparison with the usual perturbation theory}\label{sec5}

In this section, we will compare the usual perturbation theory with
our solution, reveal their consistency and relation, and point out
what is more in our solution and what is different between them.
Furthermore, we expect to reveal the features, significance and
possible applications of our solution in theory. We will
respectively investigate and discuss the cases comparing with the
time-independent, time-dependent perturbation theories and the
non-perturbative solution method.

\subsection{Comparison with the time-independent perturbation theory}

The usual time-independent (stationary) perturbation theory is
mainly to study the stationary wave equation in order to obtain the
perturbative energies and perturbative states. However, our solution
focus on the development of quantum states, which is a solution of
the Schr\"{o}dinger dynamical equation. Although their main purposes
are different at their start points, but they are consistent.
Moreover, our exact solution is also able to apply to the
calculation of transition probability, perturbed energy and
perturbed state, which will be seen clearly in Ref. \cite{My2}.
Moreover, our exact solution has been successfully applied to open
system dynamics \cite{My3}.

In order to verify that the time-independent perturbation theory is
consistent with our solution at any order approximation, we suppose
that in the initial state, the system is in the eigen state of the
total Hamiltonian, that is \beq
H\ket{\Psi_{E_T}(0)}={E_T}\ket{\Psi_{E_T}(0)}.\eeq It is clear that
\beqa \ket{\Psi_{E_T}(t)}&=&\e^{-\I {E_T}
t}\ket{\Psi_{E_T}(0)}=\sum_{\gamma,\gamma^\prime}\left\{\e^{-\I
E_{\gamma} t}\delta_{\gamma\gamma^\prime} + \sum_{l=1}^\infty
A_l^{\gamma\gamma^\prime}(t) \right\}
\left[\diracsp{\Phi^{\gamma^\prime}}{\Psi_{E_T}(0)}\right]\ket{\Phi^\gamma}.
\eeqa  Calculating the $K$th time derivative of this equation and
then setting $t=0$, we obtain \beqa \label{ktdess}{E_T}^K
\ket{\Psi_{E_T}(0)}&=&\sum_{\gamma,\gamma^\prime}\left\{E^K_{\gamma}
\delta_{\gamma\gamma^\prime} + (\I)^K\sum_{l=1}^\infty
\left.\frac{\d^K {A}_l^{\gamma\gamma^\prime}(t)}{\d t^K}
\right|_{t=0}\right\} a_{\gamma^\prime}\ket{\Phi^\gamma}\\
&=& \sum_{\gamma,\gamma^\prime}\left\{E^K_{\gamma}
\delta_{\gamma\gamma^\prime} + \sum_{l=1}^\infty
B_l^{\gamma\gamma^\prime}(K)\right\}
a_{\gamma^\prime}\ket{\Phi^\gamma},\eeqa where we have used the fact
that \beq
\ket{\Psi_{E_T}(0)}=\sum_{\gamma}a_{\gamma}\ket{\Phi^\gamma}, \quad
a_\gamma= \diracsp{\Phi^{\gamma}}{\Psi_{E_T}(0)},\eeq \beq
\left.\frac{\d^K {A}_l^{\gamma\gamma^\prime}(t)}{\d t^K}
\right|_{t=0}=(-\I)^K B_l^{\gamma\gamma^\prime}(K).\eeq It is easy
to obtain that \beqa \left.\frac{\d^K
{A}_l^{\gamma\gamma^\prime}(t)}{\d t^K}
\right|_{t=0}&=&(-\I)^K\sum_{\gamma_1,\cdots,\gamma_{l+1}}\left[
\sum_{i=1}^{l+1}(-1)^{i-1}\frac{E^K_{\gamma_i}}{d_i(E[\gamma,l])}\right]\left[
\prod_{j=1}^{l}H_1^{\gamma_j\gamma_{j+1}}\right]
\delta_{\gamma_1\gamma}\delta_{\gamma_{l+1}\gamma^\prime},\\
B_l^{\gamma\gamma^\prime}(K)&=&\sum_{\gamma_1,\cdots,\gamma_{l+1}}
C_l^K(E[\gamma,l])\theta(K-l)\left[
\prod_{j=1}^{l}H_1^{\gamma_j\gamma_{j+1}}\right]
\delta_{\gamma_1\gamma}\delta_{\gamma_{l+1}\gamma^\prime}.\label{ktdbeta}\eeqa
where we have used our identity (\ref{myi}).  This means that if
$l>K$ \beq \left.\frac{\d^K {A}_{l>K}^{\gamma\gamma^\prime}(t)}{\d
t^K} \right|_{t=0}=(-\I)^K B_l^{\gamma\gamma^\prime}(K>l)=0.\eeq In
special \beq \left.\frac{\d {A}_{l}^{\gamma\gamma^\prime}(t)}{\d t}
\right|_{t=0}=-\I B_1^{\gamma\gamma^\prime}(1)=-\I
H_1^{\gamma\gamma^\prime}\delta_{l1}.\eeq \beq \left.\frac{\d^2
{A}_{l}^{\gamma\gamma^\prime}(t)}{\d t^2}
\right|_{t=0}=-B_l^{\gamma\gamma^\prime}(2)=-
\left(E_\gamma+E_{\gamma^\prime}\right)H_1^{\gamma\gamma^\prime}\delta_{l1}
-\sum_{\gamma_1}H_1^{\gamma\gamma_1}H_1^{\gamma_1\gamma^\prime}\delta_{l2}.\eeq
In other words, the summation to $l$ in the right side of
Eq.(\ref{ktdess}) is cut off to $K$. Therefore, by left multiplying
$\bra{\Phi^\gamma}$ to Eq.(\ref{ktdess}) we obtain \beq
\label{pee}{E_T}^K a_\gamma = E^K_{\gamma}a_\gamma
 + \sum_{\gamma^\prime}\sum_{l=1}^K B_l^{\gamma\gamma^\prime}(K)
a_{\gamma^\prime}. \eeq When $K=1$, the above equation backs to the
start point of time-independent perturbation theory \beq
\label{peek1} {E_T}a_\gamma =E_\gamma a_\gamma+\sum_{\gamma^\prime}
H_1^{\gamma\gamma^\prime}a_{\gamma}^\prime . \eeq It implies that it
is consistent with our solution. In order to certify that our
solution is compatible with the time-independent perturbation
theory, we have to consider the cases when $K\geq 2$. By using
$E_T^K a_\gamma=E_T\left(E_T^{K-1} a_\gamma\right)$, substituting
Eq.(\ref{pee}) twice, then moving the second term to the right side,
we have \beqa \label{cppee} E_\gamma^{K-1}E_T a_\gamma &=&E_\gamma^K
a_\gamma+\sum_{\gamma^\prime}
\sum_{l=1}^{K}B_l^{\gamma\gamma^\prime}(K)
a_{\gamma^\prime}-\sum_{\gamma^\prime}
\sum_{l=1}^{K-1}B_l^{\gamma\gamma^\prime}(K-1)
\sum_{\gamma^{\prime\prime}}\left(E_{\gamma^\prime}
\delta_{\gamma^\prime\gamma^{\prime\prime}}+H_1^{\gamma^\prime\gamma^{\prime\prime}}\right)
a_{\gamma^{\prime\prime}}\nonumber\\
&=&E_\gamma^K a_\gamma+\sum_{\gamma^\prime}
\sum_{l=1}^{K-1}\left[B_l^{\gamma\gamma^\prime}(K)-E_{\gamma^\prime}
B_l^{\gamma\gamma^\prime}(K-1)\right]
a_{\gamma^\prime}\nonumber\\
&
&+\sum_{\gamma^\prime}\left(\prod_{j=1}^KH_1^{\gamma_j\gamma_{j+1}}\right)
\delta_{\gamma\gamma_1}\delta_{\gamma_{l+1}\gamma^\prime}a_{\gamma^\prime}
-\sum_{\gamma^\prime} \sum_{l=1}^{K-1}B_l^{\gamma\gamma^\prime}(K-1)
\sum_{\gamma^{\prime\prime}}H_1^{\gamma^\prime\gamma^{\prime\prime}}
a_{\gamma^{\prime\prime}}.\eeqa Since when $K\geq 2$\beq
\label{c1minus}
C_1^K(E[\gamma,1])-E_{\gamma_2}C_1^{K-1}(E[\gamma,1])=E_{\gamma_1}^{K-1},\eeq
and again $l\geq 2$ \beq \label{clminus}
C_l^K(E[\gamma,l])-E_{\gamma_{l+1}}C_l^{K-1}(E[\gamma,l])=C_{l-1}^{K-1}(E[\gamma,l-1]),\eeq
which has been proved in appendix A. Obviously, when $K=2$,
Eq.(\ref{cppee}) becomes \beq E_\gamma E_T a_\gamma =E_\gamma^2
a_\gamma+ E_\gamma\sum_{\gamma^\prime}
H_1^{\gamma\gamma^\prime}a_{\gamma^\prime} .\eeq It can back to
Eq.(\ref{peek1}). Similarly, when $K\geq 3$, so do they. In fact,
from Eqs.(\ref{c1minus},\ref{clminus}), Eq.(\ref{cppee}) becomes
\beqa E_\gamma^{K-1}E_T a_\gamma&=&E_\gamma^K a_\gamma
+E_\gamma^{K-1}\sum_{\gamma^\prime}H_1^{\gamma\gamma^\prime}
a_{\gamma^\prime},\eeqa where we have used the fact \beqa
\label{cppee2}& &\sum_{\gamma^\prime}
\sum_{l=2}^{K-1}\left[B_l^{\gamma\gamma^\prime}(K)-E_{\gamma^\prime}
B_l^{\gamma\gamma^\prime}(K-1)\right]
a_{\gamma^\prime}+\sum_{\gamma^\prime}\left(\prod_{j=1}^KH_1^{\gamma_j\gamma_{j+1}}\right)
\delta_{\gamma\gamma_1}\delta_{\gamma_{l+1}\gamma^\prime}a_{\gamma^\prime}
\nonumber\\
& &-\sum_{\gamma^\prime}
\sum_{l=1}^{K-1}B_l^{\gamma\gamma^\prime}(K-1)
\sum_{\gamma^{\prime\prime}}H_1^{\gamma^\prime\gamma^{\prime\prime}}
a_{\gamma^{\prime\prime}}=0.\eeqa Its proof is not difficult because
we can derive out \beqa & &
\sum_{\gamma^\prime}\sum_{l=2}^{K-1}\left[B_l^{\gamma\gamma^\prime}(K)-E_{\gamma^\prime}
B_l^{\gamma\gamma^\prime}(K-1)\right] a_{\gamma^\prime}\nonumber\\
& & \quad =\sum_{\gamma^\prime}
\sum_{l=2}^{K-1}\sum_{\gamma_1,\cdots,\gamma_{l+1}}\left[C_l^K(E[\gamma,l])-E_{\gamma^\prime}
C_l^{K-1}(E[\gamma,l])\right]\left(\prod_{j=1}^l
H_1^{\gamma_j\gamma_{j+1}}\right)\delta_{\gamma\gamma_1}\delta_{\gamma_{l+1}\gamma^\prime}
a_{\gamma^\prime}\nonumber\\
& &\quad
=\sum_{\gamma^\prime}\sum_{l=2}^{K-1}\sum_{\gamma_1,\cdots,\gamma_{l+1}}
C_{l-1}^{K-1}(E[\gamma,l-1])\left(\prod_{j=1}^l
H_1^{\gamma_j\gamma_{j+1}}\right)\delta_{\gamma\gamma_1}\delta_{\gamma_{l+1}\gamma^\prime}
a_{\gamma^\prime}\nonumber\\
& &\quad
=\sum_{\gamma^\prime}\sum_{l=1}^{K-2}\sum_{\gamma_1,\cdots,\gamma_{l+1}}
C_{l}^{K-1}(E[\gamma,l])\left(\prod_{j=1}^{l}
H_1^{\gamma_j\gamma_{j+1}}\right)H_1^{\gamma_{l+1}\gamma^\prime}
\delta_{\gamma\gamma_1} a_{\gamma^\prime}\nonumber\\
& &\quad =\sum_{\gamma^\prime,\gamma^\prime}\sum_{l=1}^{K-2}
B_l^{\gamma\gamma^\prime}(K-1)H_1^{\gamma^\prime\gamma^{\prime\prime}}
a_{\gamma^{\prime\prime}}.\eeqa The first equality has used the
definitions of $B_l^{\gamma\gamma^\prime}(K)$, the second equality
has used Eq.(\ref{clminus}), the third equality sets $l-1\rightarrow
l$ and sums over $\gamma_{l+2}$, the forth equality sets
$\gamma^\prime\rightarrow \gamma^{\prime\prime}$, inserts
$\sum_{\gamma^\prime}\delta_{\gamma_{l+1}\gamma^\prime}$ and uses
the definition of $B_l^{\gamma\gamma^\prime}(K)$ again. Substituting
it into the left side of Eq.(\ref{cppee2}) and using
$B_{K-1}^{\gamma\gamma^\prime}(K-1)$ expression, we can finish the
proof of Eq.(\ref{cppee2}). Actually, the above proof further
verifies the correctness of our solution from the usual perturbation
theory.

In our point of view, only if the expression of the high order
approximation has been obtained, can we consider its contribution in
the time-independent perturbation theory since its method is to find
a given order approximation. However, this task needs to solve a
simultaneous equation system, it will be heavy for the enough high
order approximation. In Ref. \cite{My2}, we will give a method to
find the improved forms of perturbed energy and perturbed state
absorbing the partial contributions from the high order even all
order approximations of perturbation.

\subsection{Comparison with the non-perturbative method}

Now, let us see what is more in our solution than the
nonperturbative method. Actually, since $H$ is not explicitly
time-dependent, its eigenvectors can be given formally by so-called
Lippmann-Schwinger equations as follows \cite{LSE}: \beqa
\ket{\Psi^\gamma_{\rm S}(\pm)}&=&
\ket{\Phi^\gamma}+\frac{1}{E^\gamma-H_0\pm
\I\eta}H_1\ket{\Psi^\gamma_{\rm S}(\pm)}\\
&=& \ket{\Phi^\gamma}+G_0^\gamma(\pm)H_1\ket{\Psi^\gamma_{\rm
S}(\pm)} \\
&=&\ket{\Phi^\gamma}+\frac{1}{E_\gamma-H\pm
\I\eta}H_1\ket{\Phi^\gamma}\\
&=&\ket{\Phi^\gamma}+G^\gamma(\pm)H_1\ket{\Phi^\gamma},\eeqa where
the complete Green's function $G^\gamma(\pm)=1/(E_\gamma-H\pm
\I\eta)$ and the unperturbed Green's function
$G^\gamma_0(\pm)=1/(E^\gamma-H_0\pm \I\eta)$ satisfy the Dyson's
equation \beqa
G^\gamma(\pm)&=&G_0^\gamma(\pm)+G_0^\gamma(\pm)H_1G^\gamma(\pm)\\
&=&\sum_{l=0}^\infty
\left(G_0^\gamma(\pm)H_1\right)^lG_0^\gamma(\pm). \eeqa Here, the
subscript``S" means the stationary solution: \beq H
\ket{\Psi^\gamma_{\rm S}(\pm)}=E_\gamma \ket{\Psi^\gamma_{\rm
S}(\pm)}.\eeq More strictly, we should use the ``in" and ``out"
states to express it \cite{inout}. In historical literature, this
solution is known as so-called non-perturbative one. It has played
an important role in the formal scattering theory \cite{Dyson}.

Back to our attempt, for a given initial state $\ket{\Psi(0)}$, we
have \beqa
\ket{\Psi(0)}&=&\sum_{\gamma^\prime}\diracsp{\Psi^{\gamma^\prime}_{\rm
S}(\pm)}{\Psi(0)}\ket{\Psi^{\gamma^\prime}_{\rm S}(\pm)}.\eeqa
Acting the time evolution operator on it, we immediately obtain
\beqa
\ket{\Psi(t)}&=&\sum_{\gamma^\prime}\diracsp{\Psi^{\gamma^\prime}_{\rm
S}(\pm)}{\Psi(0)}\e^{-\I
E_{\gamma^\prime}t}\ket{\Psi^{\gamma^\prime}_{\rm
S}(\pm)}\nonumber\\&=&\sum_{\gamma^\prime}
\bra{\Phi^{\gamma^\prime}}\sum_{l=0}^\infty\left(
H_1G_0^{\gamma^\prime}(\mp) \right)^l\ket{\Psi(0)}\e^{-\I
E_{\gamma^\prime}t} \sum_{l^\prime=0}^\infty
\left(G^{\gamma^\prime}(\pm)
H_1\right)^{l^\prime}\ket{\Phi^{\gamma^\prime}}\nonumber\\
&=&\sum_{\gamma,\gamma^\prime}\left[\e^{-\I E_\gamma
t}\delta_{\gamma\gamma^\prime}\diracsp{\Phi^{\gamma^\prime}}{\Psi(0)}
+\sum_{\stackrel{\scriptstyle l,l^\prime=0}{l+l^\prime\neq
0}}^\infty
\bra{\Phi^{\gamma^\prime}}\left(H_1G_0^{\gamma^\prime}(\mp)
\right)^l\ket{\Psi(0)}
\bra{\Phi^{\gamma}}\left(G_0^{\gamma^\prime}(\pm)
H_1\right)^{l^\prime}\ket{\Phi^{\gamma^\prime}}\e^{-\I
E_{\gamma^\prime} t}\right]\ket{\Phi^{\gamma}},\hskip 0.5cm\eeqa
comparing this result with our solution (\ref{ouress}), we can say
that our solution has finished the calculations of explicit form of
the expanding coefficients (matrix elements) in the second term of
the above equation using our method and rearrange the resulting
terms according with the power of elements of the perturbing
Hamiltonian matrix $H_1^{\gamma_j\gamma_{j+1}}$. This implies that
our solution can have more physical contents and more explicit
significance. Therefore, we think that our solution is a new
development of the stationary perturbation theory. Seemingly, its
physical results are consistent with the Feynman's diagram expansion
of Dyson's series, but its general form is explicit, and convenient
to calculate the time evolution of states with time. Specially, we
can obtain the improved form of perturbed solution that absorbs
partial contributions from the high order even all order
approximations in our subsequent works \cite{My2}, in which, it will
be seen that our improved scheme of perturbation theory based on our
exact solution has higher efficiency and better precision in the
calculation.

\subsection{Comparison with the time-dependent perturbation theory}

The aim of our solution is similar to the time-dependent
perturbation theory. But their methods are different. The usual
time-dependent perturbation theory \cite{diracpqm} considers a
quantum state initially in the eigenvector of the Hamiltonian $H_0$,
and a subsequent evolution of system caused by the application of an
explicitly time-dependent potential $V=\lambda v(t)$, where
$\lambda$ is small enough to indicate $V$ as a perturbation, that
is, \beq\label{seqintd} \I\frac{\partial}{\partial
t}\ket{\Psi(t)}=(H_0+\lambda v(t))\ket{\Psi(t)}\eeq with the initial
state \beq\label{inicintd} \ket{\Psi(0)}=\ket{\Phi^{\alpha}}. \eeq
It is often used to study the cases that the system returns to an
eigenvector $\ket{\Phi_f}$ of the Hamiltonian $H_0$ when the action
of the perturbing potential becomes negligible.

In order to compare it with our solution, let us first recall the
time-dependent perturbation method. Note that the state
$\ket{\Psi(t)}$ at time $t$ can be expanded by the complete
orthogonal system of the eigenvectors $\ket{\Phi^\gamma}$ of the
Hamiltonian $H_0$, that is \beq \label{tdpt0}
\ket{\Psi(t)}=\sum_\gamma c_\gamma(t)\ket{\Phi^\gamma}\eeq with \beq
c_\gamma(t)=\diracsp{\Phi^\gamma}{\Psi(t)},\eeq whereas the
evolution equation (\ref{seqintd}) will become \beq\label{Ceqintd}
\I\frac{\partial}{\partial t}c_\gamma(t)=E_\gamma
c_\gamma(t)+\lambda\sum_{\gamma^\prime}v^{\gamma\gamma^\prime}(t)c_{\gamma^\prime}(t),\eeq
where
$v^{\gamma\gamma^\prime}(t)=\bra{\Phi^\gamma}v(t)\ket{\Phi^{\gamma^\prime}}$.
Now setting \beq \label{cbr} c_\gamma(t)=b_\gamma(t)\e^{-\I E_\gamma
t}, \eeq and inserting this into (\ref{Ceqintd}) give
\beq\label{Beqintd} \I\frac{\partial}{\partial
t}b_\gamma(t)=\lambda\sum_{\gamma^\prime}\e^{\I(E_\gamma-E_{\gamma^\prime})t}
v^{\gamma\gamma^\prime}(t)b_{\gamma^\prime}(t).\eeq Then, making a
series expansion of $b_\gamma(t)$ according to the power of
$\lambda$: \beq b_\gamma(t)=\sum_{l=0}\lambda^l
b^{(l)}_\gamma(t),\eeq  and setting equal the coefficients of
$\lambda^l$ on the both sides of the equation (\ref{Beqintd}), ones
find: \beqa \I\frac{\partial}{\partial t}b^{(0)}_\gamma(t)&=&0,\\
\label{bleq} \I\frac{\partial}{\partial
t}b^{(l)}_\gamma(t)&=&\sum_{\gamma^\prime}\e^{\I(E_\gamma-E_{\gamma^\prime})t}
v^{\gamma\gamma^\prime}(t)b_{\gamma^\prime}^{(l-1)}(t), \quad (l\neq
0).\eeqa From the initial condition (\ref{inicintd}), it follows
that
 $c_\gamma(0)=b_\gamma(0)=\delta_{\gamma \alpha}=b^{(0)}_\gamma(0)$
and $b_\gamma^{(l)}(0)=0$ if $l\geq 1$. By integration over the
variable $t$ this will yield \beqa
b^{(0)}_\gamma(t)&=&\delta_{\gamma \alpha},\\
b^{(1)}_\gamma(t)&=&{-\I}\int_0^t \d
t_1\e^{\I(E_\gamma-E_{\alpha})t_1}
v^{\gamma \alpha}(t_1),\\
b^{(2)}_\gamma(t)&=&{-\I}\int_0^t \d
t_2\sum_{\gamma_2}\e^{\I(E_\gamma-E_{\gamma_2})t_2}
v^{\gamma \gamma_2}(t_2)b^{(1)}_{\gamma_2}(t_2)\nonumber\\
&=& - \sum_{\gamma_2}\int_0^t\d t_2\int_0^{t_2}\d t_1
\e^{\I(E_\gamma-E_{\gamma_2})t_2}\e^{\I(E_{\gamma_2}-E_{\alpha})t_1}
v^{\gamma\gamma_2}(t_2)v^{\gamma_2 \alpha}(t_1).\eeqa If we take $v$
independent of time, the above method in the time-dependent
perturbation theory is still valid. Thus, \beq b_\gamma^{(1)}(t)=
-\frac{1}{E_\gamma-E_\alpha}\left(\e^{-\I E_\alpha t}-\e^{-\I
E_\gamma t}\right)\e^{\I E_\gamma t}v^{\gamma \alpha},\eeq \beqa
b_\gamma^{(2)}(t)&=&
\sum_{\gamma_2}\frac{1}{(E_\gamma-E_\alpha)(E_{\gamma_2}-E_\alpha)}
\left(\e^{-\I E_\alpha t}-\e^{-\I E_\gamma t}\right) \e^{\I E_\gamma
t}v^{\gamma \gamma_2}v^{\gamma_2
\alpha}\nonumber\\
& &-\sum_{\gamma_2}
\frac{1}{(E_{\gamma_2}-E_\alpha)(E_{\gamma}-E_{\gamma_2})}
\left(\e^{-\I E_{\gamma_2} t}-\e^{-\I E_\gamma t}\right) \e^{\I
E_\gamma
t}v^{\gamma \gamma_2}v^{\gamma_2 \alpha}\nonumber\\
&=& \sum_{\gamma_2}\left[
\frac{1}{(E_{\gamma}-E_{\gamma_2})(E_{\gamma}-E_\alpha)}-
\frac{1}{(E_{\gamma}-E_{\gamma_2})(E_{\gamma_2}-E_\alpha)}\e^{-\I
E_{\gamma_2} t} \e^{\I E_\gamma
t}\right.\nonumber\\
&
&\left.+\frac{1}{(E_\gamma-E_\alpha)(E_{\gamma_2}-E_\alpha)}\e^{-\I
E_{\alpha} t} \e^{\I E_\gamma t}\right]v^{\gamma
\gamma_2}v^{\gamma_2 \alpha}.\eeqa This means that \beq
c^{(1)}_\gamma(t)=\sum_{\gamma_1\gamma_2}\sum_{i=1}^2(-1)^{i-1}\frac{\e^{-\I
E_{\gamma_i}}}{d_i(E[\gamma,1])}v^{\gamma_1\gamma_2}\delta_{\gamma\gamma_1}\delta_{\gamma_2\alpha},\eeq
\beq
c^{(2)}_\gamma(t)=\sum_{\gamma_1,\gamma_2,\gamma_3}\sum_{i=1}^3(-1)^{i-1}\frac{1}{d_i(E[\gamma,2])}\e^{-\I
E_{\gamma_i}t}\delta_{\gamma\gamma_1}\delta_{\gamma_3
\alpha}v^{\gamma_1 \gamma_2}v^{\gamma_2 \gamma_3}.\eeq Now, we use
the mathematical induction to prove \beq \label{cee}
c^{(l)}_\gamma(t)=\sum_{\gamma_1,\cdots,\gamma_{l+1}}\sum_{i=1}^{l+1}(-1)^{i-1}\frac{\e^{-\I
E_{\gamma_i}t}}{d_i(E[\gamma,l])}\left(\prod_{j=1}^l
v^{\gamma_{j}\gamma_{j+1}}\right)\delta_{\gamma\gamma_1}\delta_{\gamma_{l+1}\alpha}.\eeq
Obviously, we have seen that it is valid for $l=1,2$. Suppose it is
also valid for a given $n$, thus form Eqs.(\ref{bleq}) and
$b_\gamma^{(l)}(0)=0,\;l\geq 1$ it follows that \beq
b^{(n+1)}_\beta(t)={-\I}\int_0^t \d \tau
\sum_{\gamma}\e^{\I(E_\beta-E_{\gamma})\tau} v^{\beta
\gamma}b_\gamma^{(n)}(\tau) .\eeq Substitute Eqs.(\ref{cbr}) and
(\ref{cee}) we obtain \beqa \label{ceep2}
\!\!\!\!\!\!b^{(n+1)}_\beta(t)&=&{-\I}\int_0^t \d \tau
\sum_{\gamma}\e^{\I E_\beta\tau} v^{\beta
\gamma}c_\gamma^{(n)}(\tau)\nonumber\\
&=&{-\I}\int_0^t \d \tau \sum_{\gamma}\e^{\I E_\beta\tau} v^{\beta
\gamma}
\sum_{\gamma_1,\cdots,\gamma_{n+1}}\sum_{i=1}^{n+1}(-1)^{i-1}\frac{\e^{-\I
E_{\gamma_i}\tau}}{d_i(E[\gamma,n])}\left(\prod_{j=1}^n
v^{\gamma_{j}\gamma_{j+1}}\right)\delta_{\gamma\gamma_1}\delta_{\gamma_{n+1}\alpha}\nonumber\\
&=&\!\!\!\sum_{\gamma_1,\cdots,\gamma_{n+2}}\sum_{i=1}^{n+1}(-1)^{i-1}\frac{(\e^{-\I(
E_{\gamma_i}-E_{\gamma_{n+2}})t}-1)}{d_i(E[\gamma,n])
(E_{\gamma_i}-E_{\gamma_{n+2}})}\left(\prod_{j=1}^n
v^{\gamma_{j}\gamma_{j+1}}\right)v^{\gamma_{n+2}\gamma_1}
\delta_{\gamma_{n+2}\beta}\delta_{\gamma_{n+1}\alpha}.\eeqa In the
last equality, we have inserted
$\sum_{\gamma_{n+2}}\delta_{\gamma_{n+2}\beta}$, summed over
$\gamma$ and integral over $\tau$. In terms of the definition of
$d_i(E[\gamma,n]$, we know that
$d_i(E[\gamma,n])(E_{\gamma_i}-E_{\gamma_{n+2}})=d_i(E[\gamma,n+1])$.
Then based on our identity (\ref{myi}), we have \beq
\sum_{i=1}^{n+1}(-1)^{i-1}\frac{1}{d_i(E[\gamma,n+1])}
=-(-1)^{(n+2)-1}\frac{1}{d_{n+2}(E[\gamma,n+1])}.\eeq Thus
Eq.(\ref{ceep2}) becomes \beqa \label{ceep3}
b^{(n+1)}_\beta(t)&=&\sum_{\gamma_1,\cdots,\gamma_{n+2}}\sum_{i=1}^{n+2}(-1)^{i-1}\frac{\e^{-\I
E_{\gamma_i}t}}{d_i(E[\gamma,n+1])}\left(\prod_{j=1}^n
v^{\gamma_{j}\gamma_{j+1}}\right)v^{\gamma_{n+2}\gamma_1}
\delta_{\gamma_{n+2}\beta}\delta_{\gamma_{n+1}\alpha}\e^{\I
E_{\beta}t}.\eeqa Set the index taking turns, that is,
$\gamma_1\rightarrow\gamma_{n+2}$ and
$\gamma_i+1\rightarrow\gamma_i$,$(n+1)\geq i\geq 1)$. Again from the
definition of $d_i(E[\gamma,n]$, we can verify easily under the
above index taking turns, \beq
\sum_{i=1}^{n+2}(-1)^{i-1}\frac{\e^{-\I
E_{\gamma_i}t}}{d_i(E[\gamma,n+1])}\longrightarrow
\sum_{i=1}^{n+2}(-1)^{i-1}\frac{\e^{-\I
E_{\gamma_i}t}}{d_i(E[\gamma,n+1])}.\eeq Therefore we obtain the
conclusion \beq c_\beta^{(n+1)}(t)=b^{(n+1)}_\beta(t)\e^{-\I E_\beta
t}=\sum_{\gamma_1,\cdots,\gamma_{n+2}}\sum_{i=1}^{n+2}(-1)^{i-1}\frac{\e^{-\I
E_{\gamma_i}t}}{d_i(E[\gamma,n+1])}\left(\prod_{j=1}^{n+1}
v^{\gamma_{j}\gamma_{j+1}}\right)
\delta_{\beta\gamma_1}\delta_{\gamma_{n+2}\alpha}.\eeq This implies
that we finish the proof of expression (\ref{cee}). Substitute it
into Eq.(\ref{tdpt0}), we immediately obtain the same expression as
our solution (\ref{ouress}).

From the statement above, it can be seen that our solution actually
finishes the task to solve the recurrence equation of
$b^{l}_\gamma(t)$ by using our new method. Although the recurrence
equation of $b^{l}_\gamma(t)$ can be solved by integral in
principle, it is indeed not easy if one does not know our identity
(\ref{myi}) and relevant relations. To our knowledge, the general
term form has not ever been obtained up to now. This is an obvious
difficulty to attempt absorbing high order approximations. However,
our solution has done it, and so it will make some calculations of
perturbation theory more convenient and more accurate.

However, we pay for the price to require that $H$ is not explicitly
dependent on time. Obviously our solution is written as a symmetric
and whole form. Moreover, our method gives up, at least in the sense
of formalization, the requirement that $V$ is small enough compared
wit $H_0$, and our proof is more general and stricter. In addition,
we do not need to consider the cases that the perturbing potential
is switched at the initial and final time, but the perturbing
potential is not explicitly dependent of time. Since we obtain the
general term, our solution must have more applications, it is more
efficient and more accurate for the practical applications, because
we can selectively absorb the partial contributions from the high
order even all order approximations. This can be called as the
improved forms of perturbed solutions, which will be given in Ref.
\cite{My2}.

\section{Discussion and conclusion}\label{sec6}

First of all, we would like to point out that our general and
explicit solution (\ref{ouress}) or its particular forms
(\ref{ffs1},\ref{ffs2}) of the Schr\"{o}dinger equation in a general
quantum system independent of time is explicit and exact in form
because all order approximations of perturbation not only are
completely included but also are clearly expressed, although it is
an infinite series. Our exact solution is in $c$-number function
rather than operator form. This means that it can inherit the same
advantage as the Feynman path integral expression. Moreover, our
exact solution is a power series of the perturbing Hamiltonian like
as the Dyson series in the interaction picture. This implies that
the cut-off approximation of perturbation can be made based on the
needed precision of the problems. It must be emphasized that when
applying our exact solution to a concrete quantum system, all we
need to do for our exact solution is the calculations of perturbing
Hamiltonian matrix and the coefficient function limitations. Here,
the ``perturbing Hamiltonian matrix" refers to the representation
matrix of the perturbing Hamiltonian in the unperturbed Hamiltonian
representation. Thus, such calculations are definitely much easier
than the calculations of path integral and Dyson series for the most
systems. On the other hand, because the general term is explicitly
given in our exact solution series, we can more conveniently
consider the contributions from the high order even all order
approximations in order to improve the precision than doing the same
thing via the Dyson series.

Although we have obtained the exact solution in general quantum
systems, this does not mean that the perturbation theory is
unnecessary because our exact solution is still an infinite power
series of perturbation. Our solution is called ``exact one" in the
sense including the contributions from all order approximations of
the perturbing Hamiltonian. In practise, if we do not intend to
apply our solution to investigations of the formal theory of quantum
mechanics, we need to cut off our exact solution series to some
given order approximation in the calculations of concrete problems.
Perhaps, one argues that our exact solution will back to the usual
perturbation theory, and it is, at most, an explicit form that can
bring out the efficiency amelioration. Nevertheless, the case is not
so. Such a view, in fact, ignores the significance of the general
term in an infinite series, and forgets the technologies to deal
with an infinite series in the present mathematics and physics. From
our point of view, since the general term is known, we can
systematically and reasonably absorb the partial contributions from
some high order even all order approximations just like one has done
in quantum field theory via summation over a series of different
order but similar feature Feynman figures. In our paper \cite{My2},
we will fixed this problem via proposing and developing an improved
scheme of perturbation theory.

In order to account for what is more in our solution, and reveal the
relations and differences between our solution and the existed
method, we compare our solution with the usual perturbation theory.
We find their consistency. In fact, our solution has finished the
task to calculate the expanding coefficients of final state in $H_0$
representation and obtain the general term up to any order using our
own method. But the usual perturbation theory only carries out this
task from some given order approximation to the next order
approximation step by step. In a sentence, more explicit and general
feature of our exact solution can lead to more physical conclusions
and applications. In addition, by comparison with the existed
perturbation theory, we actually verify the validity of our exact
solution.

It is worth pointing out that our solution, different from the
time-dependent perturbation theory, presents the explicit solution
of recurrence equation of expanding coefficients of final state in
$H_0$ representation, but we pay for the price that $H$ is not
explicitly dependent on the time. In short, there is gain and there
is lose.

It is worthy noticing that there are the apparent divergences in the
expression of our exact solution, in special, when the degeneracy
happens. These apparent divergences can be easily eliminated via the
limitation calculations. The related discussions are arranged in our
paper \cite{My2}.

Therefore, we can say that our exact solution is more explicit than
the usual non-perturbed solution of the Schr\"{o}dinger equation
because the general term is given in the $c$-number function form,
and we can say our solution is more general than the usual
perturbation theory because our deducing methods give up some
preconditions used in the usual scheme. However, our exact solution
can not be used to the time-dependent system at present. Just as its
explicit and general features, our exact solution not only has the
mathematical delicateness, but also can contain more physical
content, obtain the more efficiency and higher precision and result
in new applications, which have been revealed in our serial studies
\cite{My2,My3}.

Our physical idea is a combination of Feynman path integral spirit
and Dyson series kernel. Hence, we first find an general and
explicit expression of the time evolution operator that not only is
a $c$-number function but also a power series of perturbation
including all order approximations. From our point of view, it is
one of the most main results in our method. It must be emphasized
that the study on the time evolution operator plays a central role
in quantum dynamics and perturbation theory. Because the universal
significance of our general and explicit expression of time
evolution operator, we wish that it will have more applications in
quantum theory. Besides the perturbation theory and open system
dynamics, it is more interesting to apply our exact solution to the
other formalization study of quantum dynamics in order to further
and more powerfully show the advantages of our exact solution.

In the process of deducing our exact solution, we obtain an
expansion formula of operator binomials power. At our knowledge,
this formula is first proposed and strictly proved. Besides its
theoretical value in mathematics, we are sure it is interesting and
important for expressing some useful operator formula in quantum
physics. In addition, we prove an identity of fraction function. It
should be interesting in mathematics. Perhaps, it has other
applications to be expected finding.

We can see that in our solution, every expanding coefficient
(amplitude) before the basis vector $\ket{\Phi^\gamma}$ has some
closed time evolution factors with exponential form, and includes
all order approximations so that we can clearly understand the
dynamical behavior of quantum systems. Although our solution is
obtained in $H_0$ representation, its form in other representations
can be given by the representation transformation and the
development factors with time $t$ in the expanding coefficients
(amplitude) do not change.

Because we obtain the general term form of time evolution of quantum
state, it provides the probability considering the partial
contributions from the high order even all of order approximations.
For example, we obtain the revised Fermi's Golden Rule \cite{Fermi}.
This can be seen in our recent work \cite{My2}.

Our general and explicit solution can be successfully applied to
open systems to obtain the  master equation and exact solution. For
example, we obtain the Redfield master equation \cite{Redfield}
without using Born-Markov approximation. This can be seen in our
recent work \cite{My3}.

From the features of our solution, we believe that it will have
interesting applications in the calculation of entanglement dynamics
and decoherence process as well as the other physical quantities
dependent on the expanding coefficients. Of course, our solution is
an exact one, its advantages and features can not be fully revealed
only via the perturbative method.

In summary, our results can be thought of as theoretical
developments of quantum dynamics, and are helpful for understanding
the dynamical behavior and related subjects of general quantum
systems in both theory and application. Together with our
perturbation theory \cite{My2} and open system dynamics \cite{My3}
they can finally form the foundation of theoretical formulism of
quantum mechanics in general quantum systems. Further study on
quantum mechanics of general quantum systems is on progressing.

\section*{Acknowledgments}

We are grateful all the collaborators of our quantum theory group in
the Institute for Theoretical Physics of our university. This work
was funded by the National Fundamental Research Program of China
under No. 2001CB309310, and partially supported by the National
Natural Science Foundation of China under Grant No. 60573008.

\begin{appendix}
\renewcommand{\theequation}{\thesection\arabic{equation}}

\section{The proof of our identity}

In this appendix, we would like to prove our identity (\ref{myi}).
For simplicity in notation and universality, we replace the
variables $E_{\gamma_i}$ by $x_i$ as well as $E[\gamma,l]$ by
$x[l]$. It is clear that the common denominator $D(x[n])$ in the
above expression (\ref{myi}) (the index $l$ is replaced by $n$)
reads \beq D(x[n])=\prod_{i=1}^n\left[\prod_{j=i}^n
\left(x_i-x_{j+1}\right)\right], \eeq while the $i$-th numerator is
\beq n_i(x[n],K)=\frac{D(x[n])}{d_i(x[n])}x_i^K, \eeq and the total
numerator $N(x[n],K)$ is \beq
N(x[n],K]=\sum_{i=1}^{n+1}n_i(x[n],K).\eeq In order to simplify our
notation, we denote $n_i(x[n])=n_i(x[n],0)$. Again, introducing a
new vector \beqa x^D_1[n]&=&\{x_2,x_3,\cdots,x_n,x_{n+1}\},\\
x^D_i[n]&=&\{x_1,\cdots,x_{i-1},x_{i+1},\cdots,x_{n+1}\},\\
x^D_{n+1}[n]&=&\{x_1,x_2,\cdots,x_{n-1},x_{n}\}. \eeqa Obviously,
$x^D_i(x[n])$ with $n$ components is obtained by deleting the $i$-th
component from $x[n]$. From the definition of $n_i(x[n])$, it
follows that \beq\label{nDr} n_i(x[n])=D(x^D_i[n]). \eeq

Without loss of generality, for an arbitrary given $i$, we always
can rewrite $x^D_i[n]=y[n-1]=\{y_1,y_2,\cdots,y_{n-1},y_n\}$ and
consider the general expression of $D(y[n-1])$. It is easy to verify
that \beqa \!\!\!\!\!\!\!D(y[1])\!\!\!&=&\!\!\!-y_2n_1(y[1])+y_1n_2(y[1]), \\
 \!\!\!\!\!\!\!D(y[2])\!\!\!&=&\!\!\!y_2y_3n_1(y[2])-
y_1y_3n_2(y[2])+y_1y_2n_3(y[2]). \eeqa Thus, by mathematical
induction, we first assume that for $n\geq 2$, \beq \label{Dform}
D(y[n-1])=\sum_{i=1}^n (-1)^{(i-1)+(n-1)}p_i(y[n-1])n_i(y[n-1]),\eeq
where we have defined \beq p_i(y[n-1])=\prod_{\stackrel{\scriptstyle
j=1}{j\neq i}}^{n} y_j.\eeq As above, we have verified that the
expression (\ref{Dform}) is valid for $n=2,3$. Then, we need to
prove that the following expression \beq \label{Dformp}
D(y[n])=\sum_{i=1}^{n+1} (-1)^{(i-1)+n}p_i(y[n])n_i(y[n])\eeq is
correct.

To our purpose, we start from the proof of a conclusion of the
precondition (\ref{Dform}) as the following: \beq \label{nsum}
\sum_{i=1}^{n+1}(-1)^{i-1}n_i(y[n])=0.\eeq According to the relation
(\ref{nDr}) and substituting the precondition (\ref{Dform}), we have
\beqa
\sum_{i=1}^{n+1}(-1)^{i-1}n_i(y[n])&=&\sum_{i=1}^{n+1}(-1)^{i-1}D(y_i^D[n])
=\sum_{i=1}^{n+1}(-1)^{i-1}\left[\sum_{j=1}^{n}(-1)^{(j-1)+(n-1)}
p_j(y^D_i[n])n_j(y^D_i[n])\right]\nonumber \\
&=& (-1)^{n-1} \sum_{j=1}^n \sum_{i=1}^{n+1}(-1)^{i+j}
p_j(y^D_i[n])n_j(y^D_i[n])\label{nsum1}.\eeqa Based on the
definitions of $p_i(x[n])$ and $n_i(x[n])$, we find that \beqa
p_j(y^D_i[n])&=&\left\{
\begin{array}{ll} p_{i-1}(y^D_j[n]) &\quad \mbox{(If $i>j$)}\\[8pt]
p_i(y^D_{j+1}[n])&\quad \mbox{(If $i\leq j$)}\end{array}\right.,\\
n_j(y^D_i[n])&=&\left\{
\begin{array}{ll} n_{i-1}(y^D_j[n]) &\quad \mbox{(If $i>j$)}\\[8pt]
n_i(y^D_{j+1}[n])&\quad \mbox{(If $i\leq j$)}\end{array}\right..
\eeqa Thus, the right side of Eq.(\ref{nsum1}) becomes \beqa &
&(-1)^{n-1} \sum_{j=1}^n \sum_{i=1}^{n+1}(-1)^{i+j}
p_j(y^D_i[n])n_j(y^D_i[n])\nonumber\\
&=&(-1)^{n-1}\sum_{j=1}^n\sum_{i=j+1}^{n+1}(-1)^{i+j}p_j(y^D_i[n])n_j(y^D_i[n])
\nonumber\\& & +
(-1)^{n-1}\sum_{j=1}^n\sum_{i=1}^{j}(-1)^{i+j}p_j(y^D_i[n])n_j(y^D_i[n])\nonumber\\
&=&(-1)^{n-1}\sum_{j=1}^n\sum_{i=j+1}^{n+1}(-1)^{i+j}p_{i-1}(y^D_j[n])n_{i-1}(y^D_j[n])
\nonumber\\ & &+
(-1)^{n-1}\sum_{j=1}^n\sum_{i=1}^{j}(-1)^{i+j}p_i(y^D_{j+1}[n])n_i(y^D_{j+1}[n]).
\eeqa Setting that $i-1\rightarrow i$ for the first term and
$j+1\rightarrow j$ for the second term in the right side of the
above equation, we obtain \beqa & &(-1)^{n-1} \sum_{j=1}^n
\sum_{i=1}^{n+1}(-1)^{i+j}
p_j(y^D_i[n])n_j(y^D_i[n])\nonumber \\
&=&(-1)^{n-1}\sum_{j=1}^n\sum_{i=j}^{n}(-1)^{i+j-1}p_{i}(y^D_j[n])n_{i}(y^D_j[n])
\nonumber\\ & &+
(-1)^{n-1}\sum_{j=2}^{n+1}\sum_{i=1}^{j-1}(-1)^{i+j-1}p_i(y^D_{j}[n])n_i(y^D_{j}[n])\nonumber \\
&=&(-1)^{n}\sum_{j=1}^n\sum_{i=1}^{n}(-1)^{i+j}p_{i}(y^D_j[n])n_{i}(y^D_j[n])
\nonumber\\ & &+
(-1)^{n}\sum_{i=1}^{n+1}(-1)^{i+(n+1)}p_i(y^D_{n+1}[n])n_i(y^D_{n+1}[n]).\eeqa
Again, setting that $i\leftrightarrow j$ in the right side of the
above equation gives \beqa & & (-1)^{n-1} \sum_{j=1}^n
\sum_{i=1}^{n+1}(-1)^{i+j} p_j(y^D_i[n])n_j(y^D_i[n])\nonumber \\ &
&= (-1)^{n} \sum_{j=1}^n \sum_{i=1}^{n+1}(-1)^{i+j}
p_j(y^D_i[n])n_j(y^D_i[n]). \eeqa Thus, it implies that \beq
(-1)^{n-1} \sum_{j=1}^n \sum_{i=1}^{n+1}(-1)^{i+j}
p_j(y^D_i[n])n_j(y^D_i[n])= 0. \eeq From the relation (\ref{nsum1})
it follows that the identity (\ref{nsum}) is valid.

Now, let us back to the proof of the expression (\ref{Dformp}).
Since the definition of $D(y[n])$, we can rewrite it as \beq
D(y[n])=D(y[n-1])d_{n+1}(y[n]). \eeq Substituting our precondition
(\ref{Dform}) yields \beqa
D(y[n])&=&\sum_{i=1}^n(-1)^{(i-1)+(n-1)}p_i(y[n-1])n_i(y[n-1])d_{n+1}(y[n]).\eeqa
Note that \beqa n_i(y[n-1])d_{n+1}(y[n])&=&n_i(y[n])(y_i-y_{n+1}),\\
p_i(y[n-1])y_{n+1}&=&p_i(y[n]),\\
p_i(y[n-1])y_{i}&=&p_{n+1}(y[n]),\eeqa we have that
\beqa\label{Dformm}
\!\!\!D(y[n])\!\!\!&=&\!\!\!\sum_{i=1}^n(-1)^{(i-1)+n}p_i(y[n])n_i(y[n])
+\sum_{i=1}^n(-1)^{(i-1)+(n-1)}p_{n+1}(y[n])n_i(y[n])\nonumber\\
&=& \sum_{i=1}^n (-1)^{(i-1)+n}p_i(y[n])n_i(y[n])
+(-1)^{n-1}p_{n+1}(y[n])\left[\sum_{i=1}^n(-1)^{i-1}n_i(y[n])\right]\nonumber\\
&=& \sum_{i=1}^n (-1)^{(i-1)+n}p_i(y[n])n_i(y[n])
+(-1)^{n-1}p_{n+1}(y[n])\left[-(-1)^{n}n_{n+1}(y[n])\right]. \eeqa
In the last equality we have used the conclusion (\ref{nsum}) of our
precondition, that is \beq \sum_{i=1}^n(-1)^{i-1}n_i(y[n])=-(-1)^n
n_{n+1}(y[n]).\eeq Thus, Eq.(\ref{Dformm}) becomes \beq
D(y[n])=\sum_{i=1}^{n+1}(-1)^{(i-1)+n}p_i(y[n])n_i(y[n]).\eeq The
needed expression (\ref{Dformp}) is proved by mathematical
induction. That is, we have proved that for any $n\geq 1$, the
expression (\ref{Dformp}) is valid.

Since our proof of the conclusion (\ref{nsum}) of our precondition
is independent of $n$ ($n\geq 1$), we can, in the same way, prove
that the identity (\ref{nsum}) is correct for any $n\geq 1$.

Now, let us prove our identity (\ref{myi}). Obviously, when $K=0$ we
have \beq\label{cnzero}
\sum_{i=1}^{n+1}(-1)^{i-1}\frac{1}{d_i(x[n])}=\frac{1}{D(x[n])}
\left[\sum_{i=1}^{n+1}(-1)^{i-1}n_i(x[n])\right]=0,\eeq where we
have used the fact the identity (\ref{nsum}) is valid for any $n\geq
1$. Furthermore, we extend the definition domain of $C^K_n(x[n])$
from $K\geq n$ to $K\geq 0$, and still write its form as \beq
C^K_n(x[n])= \sum_{i=1}^{n+1}(-1)^{i-1}\frac{x_i^K}{d_i(x[n])}.\eeq
Obviously, Eq.(\ref{cnzero}) means \beq C_n^0(x[n])=0,\quad (n\geq
1).\eeq In order to consider the cases when $K\neq 0$, by using of
$d_i(x[n])(x_i-x_{n+2})=d_i(x[n+1])$ ($i\leq n+1$), we obtain\beqa
C^K_n(x[n])&=& \sum_{i=1}^{n+1}(-1)^{i-1}\frac{x_i^K
}{d_i(x[n])}\nonumber\\&=&\sum_{i=1}^{n+1}(-1)^{i-1}\frac{x_i^K
}{d_i(x[n])}\left(\frac{x_i-x_{n+2}}{x_i-x_{n+2}}\right)\nonumber\\
&=&\sum_{i=1}^{n+1}(-1)^{i-1}\frac{x_i^{K+1}
}{d_i(x[n+1])}-x_{n+2}\sum_{i=1}^{n+1}(-1)^{i-1}\frac{x_i^K
}{d_i(x[n+1])}\nonumber\\ &=&
C^{K+1}_{n+1}(x[n+1])-(-1)^{(n+2)-1}\frac{x_{n+2}^{K+1}
}{d_{n+2}(x[n+1])}-x_{n+2}\sum_{i=1}^{n+1}(-1)^{i-1}\frac{x_i^K
}{d_i(x[n+1])}. \eeqa It follows the recurrence equation as the
following \beq\label{myirq}
C^K_n(x[n])=C^{K+1}_{n+1}(x[n+1])-x_{n+2}C^{K}_{n+1}(x[n+1]).\eeq It
implies that since $C_{n}^0(x[n])=0$ for any $n\geq 1$, then
$C_{n+1}^1(x[n+1])=0$ for any $n\geq 1$ or $C_{n}^1(x[n])=0$ for any
$n\geq 2$; since $C_{n}^1(x[n])=0$ for any $n\geq 2$, then
$C_{n+1}^2(x[n+1])=0$ for any $n\geq 2$ or $C_{n}^2(x[n])=0$ for any
$n\geq 3$; $\cdots$, since $C_{n}^k(x[n])=0$ for any $n\geq (k+1)$
($k\geq 0$), then $C_{n+1}^{k+1}(x[n+1])=0$ for any $n\geq (k+1)$ or
$C_{n}^{k+1}(x[n])=0$ for any $n\geq (k+2)$; $\cdots$. In fact, the
mathematical induction tells us this result. Obviously \beq
\label{myig1} C_n^K(x[n])=0,\quad (\mbox{If $0\leq K<n$}). \eeq
Taking $K=n$ in Eq.(\ref{myirq}) and using Eq.(\ref{myig1}), we have
 \beq C_n^n(x[n])=C_{n+1}^{n+1}(x[n+1])=1, \quad (n\geq 1),\eeq where we
have used the fact that $C_1^1(x[1])=1$. Therefore, the proof of our
identity (\ref{myi}) is finally finished.

\end{appendix}



\end{document}